**Original Manuscript**

[**What shapes climate change perceptions in Africa? A random forest approach**]

[Juan B González[1*], Alfonso Sanchez[2]]

[**1** Department of Economics, Universidad Loyola, Dos Hermanas, Spain.

**2** Department of International Studies, Universidad Loyola, Dos Hermanas, Spain.]

Corresponding author: Juan B González

* juanbgonzalezblanco@gmail.com

**Abstract**

Climate change perceptions are fundamental for adaptation and environmental policy support. Although Africa is one of the most vulnerable regions to climate change, little research has focused on how climate change is perceived in the continent. Using random forest methodology, we analyse Afrobarometer data (N = 45,732), joint with climatic data, to explore what shapes climate change perceptions in Africa. We include 5 different dimensions of climate change perceptions: awareness, belief in its human cause, risk perception, need to stop it and self-efficacy. Results indicate that perceived agriculture conditions are crucial for perceiving climate change. Country-level factors and long-term changes in local weather conditions are among the most important predictors. Moreover, education level, access to information, poverty, authoritarian values, and trust in institutions shape individual climate change perceptions. Demographic effects -including religion- seem negligible. These findings suggest policymakers and environmental communicators how to frame climate change in Africa to raise awareness, gather public support and induce adaptation.

# Introduction

*The importance of climate change perceptions*

Current projections assess that climate change (CC) will likely have "severe, irreversible and pervasive impacts for people and ecosystems" [1]. Urgent mitigation and adaptation strategies are needed at both the social and individual level. However, these measures are not being implemented rapidly enough. Apart from material and institutional constraints, some cognitive barriers hinder adaptation [2,3]. Among those cognitive barriers, climate change perceptions (CCP) stand out [4]. Therefore, it is fundamental to understand how individuals perceive climate change in order to induce behavioural changes and gather policy support [5,6].

Despite the strong scientific consensus on the existence and projected impacts of CC, a relevant fraction of the public deny its existence, underestimate its risks or believe it is a natural process that cannot be stopped [7]. In contrast with the scientific consensus, there is a huge variance among individual and public CCP. Previous research has established a wide array of factors that explain this divergence [8–10].

*What shapes CCP in the Global North*

Reasonably, having access to information is fundamental for being aware of CC and its causes [11]. Likewise, higher levels of education have been associated with greater awareness and concern [12,13]. However, other studies find that information, scientific literacy or education level are not correlated to CCP [14–16]. Some factors mediate how people access, process and assimilate information. One of those factors is ideology, which influences CCP in various ways. Ideology affects the choice of information sources, as individuals seek those that match and reinforce their previous beliefs [17,18]. But even when they are presented with the same information, it is processed differently depending on ideology. For instance, Hart and Nisbet [16] presented the same piece about possible CC impacts on health to US Democrats and Republicans. It increased CC risk perceptions and policy support among Democrats, but it resulted in a "boomerang effect" among Republicans, who left more convinced of their previous scepticism. This has been explained in terms of ideologically motivated reasoning [15]. To reduce cognitive dissonance and peer pressure, novel information about CC is processed and assimilated to match previous beliefs, even when they conflict. Thus, ideology shapes how individuals perceive climate change [12,19].

Given the statistical nature of CC and the psychological distance to it, information about it does not usually elicit strong emotional responses, hindering its perception [9,20,21]. In contrast, personal experience of local weather is sensually and emotionally salient, so it can irrationally substitute rigorous but abstract scientific information [3,22–24]. Personal experience of extreme climate events such as hurricanes, floods or extreme temperatures increases CCP [25–27]. But the effect of personal experience also holds for short- and long-term temperature anomalies [24,28–31]. Therefore, personal experience of local weather is another factor that influences CCP.

Information about CC might be accessed, processed, and assimilated through biased processes, but it can also be directly ignored. The psychological distance of CC contrasts with daily material concerns, relegating CC to irrelevance. In other words, people may have a "finite pool of worry" which may be full of more immediate concerns than CC, limiting its perception [8]. In line with this hypothesis, GDP and GDP growth, unemployment and household income have been related to CCP [11,13,32–34].

Finally, religion has been found to have a significant influence on CCP. Like political ideology, religion can push people to group-thinking and motivated reasoning. For instance, an individual who believes in an almighty God is more likely to attribute CC to God's will rather than to human activity. Attending religious services and having religious beliefs have both been found to affect CCP [30,35–38]. Apart from religion, other demographics such as gender, race or age do not have consistent effects on CCP [10].

*CCP in Africa*

The projected impacts of climate change are unevenly distributed across regions, and Africa will be among the most affected [1]. Although adaptation is especially urgent in Africa, little research has focused on the cognitive barriers to adaptation in the continent. For instance, just 3% of studies meta-analysed by Van Valkengoed and Steg [4] and 1.7% of those meta-analysed by Hornsey et al. [10] were conducted in Africa. As predictors of CCP vary widely across regions [11,13], the applicability of non-African research is, at least, questionable.

Research on CCP in Africa is scarce. Beyond the local case-study level [32,39], there are few cross-African studies [11,13]. Moreover, these studies rely on the same surveys (Gallup Poll 2007-2010) and, due to data constraints, just include a handful of CCP predictors. For instance, significant predictors such as ideology or local weather changes are not included in the analysis. Building upon that research, this study explicitly addresses what shapes climate change perceptions in Africa. The topic is approached holistically, as the importance of education, access to information, ideology, experience of local weather, religion, demographics, and economic variables (among others) to predict individual CCP are assessed simultaneously. This analysis offers a clear picture of how CCP are constructed in Africa.

**Materials and Methods**

*Datasets*

For extracting CCP variables and most predictors, we use the $7^{th}$ round of the Afrobarometer [7], conducted between 2016 and 2018. It comprises more than 45,000 observations from 34 African countries. Except for some small countries, it is georeferenced at the second administrative level (see SI 1). This allows a high resolution for relating CCP to climatic variables, a link unstudied beyond the first administrative level across Africa.

For constructing local weather variables, we use two different datasets. First, we obtain monthly precipitation, maximum and mean temperature for the period 1961-2019 from the CRU 4.0 dataset [40]. Second, we complement those variables with the standardized precipitation evapotranspiration index (SPEI), which robustly measures drought, from the SPEI 2.6 database [41]. Both datasets offer a spatial resolution of 0.5° x 0.5°.

*Measures*

CCP are the dependent variables in this study. Specifically, we include the following CCP variables. *CC awareness* accounts for the question "Have you heard about climate change or haven't you had the chance to hear about this yet?" (1=yes, 0=no). Those aware were asked the following questions. *Human cause*: "People have different ideas about what causes climate change. What about you, which of the following do you think is the main cause of climate change, or haven't you heard enough to say?" (1=human cause, 0=other). *CC risk perception*: "Do you think climate change is making life in [your country] better or worse, or haven't you heard enough to say?" (0=better/the same, 1=worse). *Need to stop CC*: "Do you think that climate change needs to be stopped?" (0-1). Those who believed CC needs to be stopped were asked a last question. *Self-efficacy*: "How much do you think that ordinary [citizens] can do to stop climate change?" (1=a little bit/a lot, 0=nothing).

For climatic variables, we first superpose the CRU and SPEI data grids on the GADM second administrative level map of Africa. As some administrative areas intersect with more than one pixel, we aggregate their values using two alternative functions: mean and maximum. Further analysis is made with mean values, but it is also robust to the use of maximum values across grids. Second, we compute the long-term anomalies for temperature and precipitation data (SPEI is already standardised against a long-term baseline). We use annual values (the year before the individual was surveyed) standardised against the 1961-1990 baseline.

Additionally, 67 potential correlates to CCP are extracted from the Afrobarometer. They account for ideology, economic conditions, demographics, access to information, education, intention to migrate or agricultural perceptions, among others. As collinearity can bias variable importance measures in Random Forest [42], we conduct a Principal Component Analysis. This creates orthogonal linear combinations of highly correlated variables, reducing thus the problem of collinearity. Only combinations with a Cronbach's alpha higher than 0.7 are kept, a conservative level often used in the field [14,38]. Adding Afrobarometer and climatic variables, we have 51 potential predictors of CCP. For all independent variables, missing values are handled using non-parametric imputation with the R package *missRanger* [43].

*Methods*

We analyse what shapes climate change perceptions in Africa using Random Forest methodology [44]. This machine-learning approach uses non-parametric recursive partitioning to produce models with high predictive accuracy [11]. It can handle high-dimensional (with a large number of predictors) multilevel datasets with high-level interactions and non-linear relations [42], so it is ideal for our dataset. For each dependent variable, we grow a random forest composed of 1,000 trees with a minimum node size of 5, using the *ranger* package in R [45]

Despite its advantages, Random Forests are not easily interpretable on their own. To interpret them, we use some additional measures. First, we compute the variable importance measure, that ranks predictors by their predictive power (including direct and indirect effects on the dependent variable). We use the corrected Gini method to do so, because it shows no bias towards predictors with more classes, in contrast to the permutation method [46]. This measure shows which are the most important predictors that shape CCP but does not assess whether they are significant or not.

Second, we use the Altmann permutation method to compute p-values and test predictor significance, using 100 permutations for each forest, as recommended by the authors [47]. We use the *ranger* package in R [45] to compute those measures. Finally, we use partial dependence plots to illustrate the magnitude and direction of the direct effects of significant predictors. Partial dependence plots work like marginal effects in logistic regression models: they predict responses for each level of the predictor while holding the rest of the variables constant. We use the *randomForestSRC* package for generating these plots [48].

**Results**

*CC awareness*

Fig 1 presents the most important predictors for being aware of climate change in Africa. Education level and the frequency of access to online news (via the internet and social media) are fundamental for CC awareness. Both have positive effects. Perceiving that climate conditions for agricultural production (*agric. cond.*) have changed in the last decade is positively related to CC awareness, but the effect is higher for perceived worsening (positive values), as Fig 1B illustrates. Ideology and interest in politics are also important covariates. Authoritarians (being favourable to one-party rule) significantly decreases awareness, while talking about politics has the opposite effect. We find a gender gap for awareness, as women are about 5.4% less likely to know about climate change. Long-term changes in weather conditions at the second administrative level are important predictors of being aware of CC. Higher temperatures, less rainfall, and more severe droughts (SPEI) are associated with higher CC awareness, but their direct effects are of less magnitude than education, information, or ideology. Speaking a western language has a significant impact of almost +4% on CC awareness. Regarding religion, we find mixed results: while being religious (any denomination) has a positive relation, supporting the rule of religious law reduces CC awareness.

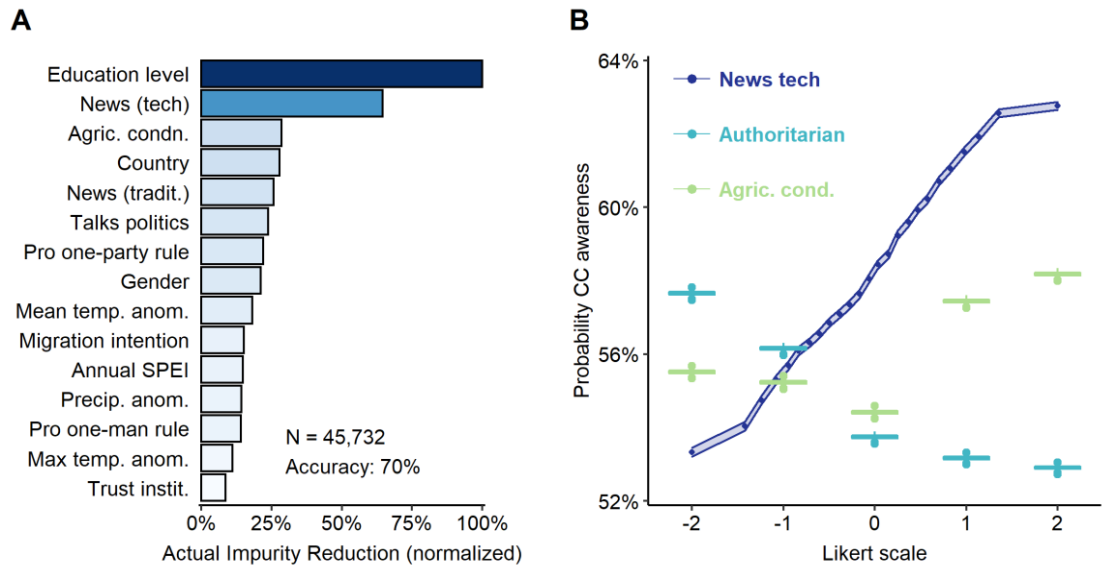

**Fig 1. Key predictors of climate change awareness.**

(A) Top 15 predictors of CC awareness. (B) Partial dependence plot of direct effects of access to online news (*news tech*), being favourable to one-party rule (*authoritarian*) and perceived agricultural conditions (*agric. cond.*)

*Belief in human cause*

Fig 2 shows the most important predictors of believing in the human causation of CC. Local weather changes are the main predictors, above education, access to information or ideology. A 1 SD rise of mean temperatures is associated with almost a 6% increase in the belief that CC is caused by human action, as illustrated in Fig 2B. Changing precipitations have the opposite effect but with less magnitude. Education level and access to online information also have important positive effects over the belief in the human causation of CC. Trust in institutions and authoritarian and

intolerant —towards other religions, ethnic groups, and nationalities— ideologies are associated with reduced belief. Agricultural and drought perceptions remain important for predicting this dimension of CCP. Finally, being religious is insignificant for predicting the belief in the anthropogenic nature of CC.

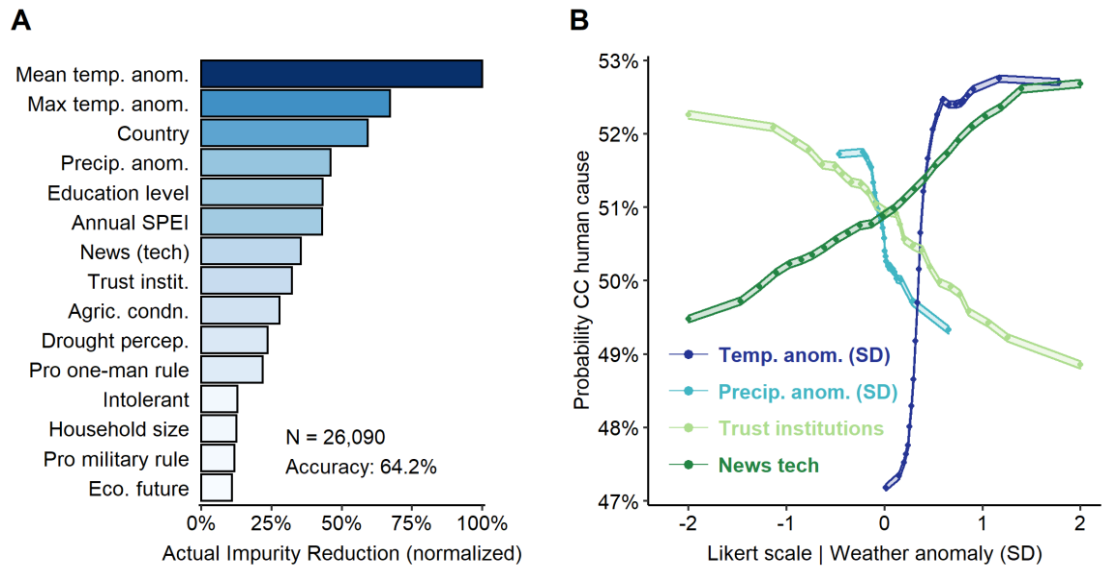

**Fig 2. Key predictors of belief in human causation of climate change.**

(A) Top 15 predictors of belief in human causation of CC. (B) Partial dependence plot of direct effects of mean temperature anomalies (*temp. anom.*), precipitation anomalies (*precip. anom.*), trust in institutions (*trust* institutions) and access to online news (*news tech*).

*CC risk perception*

The key predictors of CC risk perceptions are presented in Fig 3. Perceived agricultural conditions, followed by drought perception, are crucial for assessing the risks CC poses to citizens in Africa. Those who perceive better agricultural conditions are less likely to consider CC as a risk than those who perceive no changes, whereas those who perceive worse conditions are significantly more likely. Believing in the human cause of CC is positively related to risk perceptions. Local weather changes maintain their importance. Temperature anomalies have a positive effect, while the effect is the opposite for precipitation. Authoritarianism —supporting one-man, one-party or military rule, is negatively related to perceived risks. Again, trust in institutions is associated with reduced risk perception. Poverty, on the other hand, shows the opposite direction: households with fewer resources perceive greater risks from CC than wealthier ones, both in urban and rural areas. Finally, speaking a western language reduces risk perception by almost 6%.

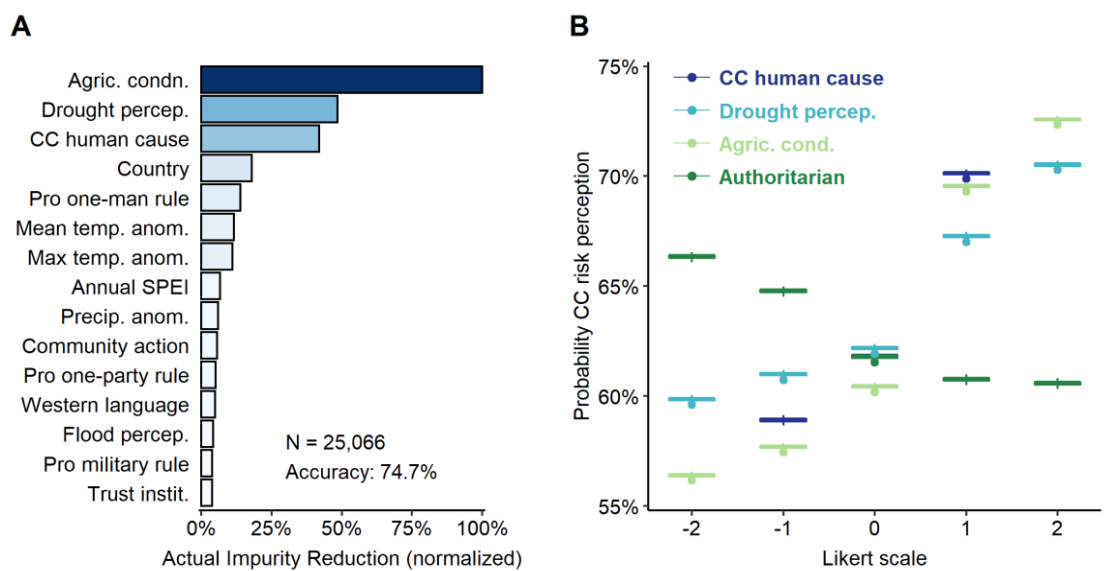

**Fig 3. Key predictors of climate change risk perception.**

(A) Top 15 predictors of CC risk perception. (B) Partial dependence plot of direct effects of belief in human causation of CC (*CC human cause*), perceived severity of droughts (*drought percep.*), perceived agricultural conditions (*agric. cond.*) and being favourable to one-man rule (*authoritarian*).

*Need to stop CC*

As Fig 4 shows, CC risk perceptions and the belief in human causation of CC are the top predictors of the need to stop it. Both have important positive impacts on the belief that action is needed, with a maximum effect of more than 15% for risk perception and 12% for human cause. Like other dimensions of CCP, support for action against CC is positively related to worse agricultural conditions and higher temperature anomalies, and negatively to rainfall anomalies. Individuals with fewer resources and with democratic values are more convinced of the need to stop CC, whereas authoritarians and western languages speakers are less convinced. Perceived corruption has a non-monotonous effect. Both higher and lower perceived corruption levels lead to decreased action support, while moderate perceptions lead to higher probabilities.

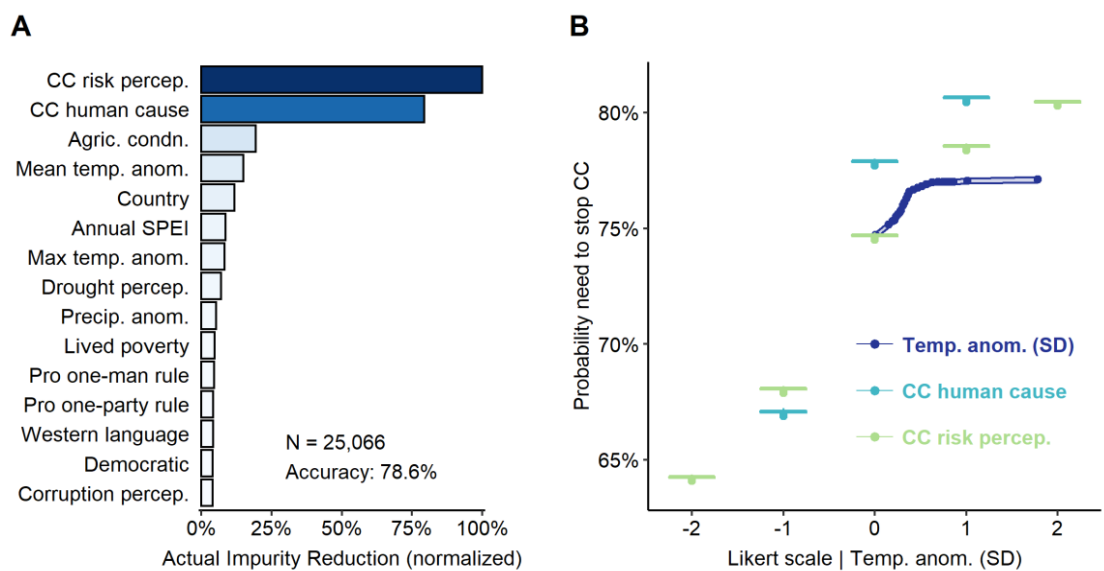

**Fig 4. Key predictors of believing climate change must be stopped.**

(A) Top 15 predictors of need to stop CC. (B) Partial dependence plot of direct effects of mean temperature anomalies (*temp. anom.*), belief in human causation of CC (*CC human cause*), and perceived risk from CC (*CC risk percep.*) on the need to stop CC.

*Self-efficacy*

Fig 5 present the key predictors of self-efficacy — the perceived effectiveness of ordinary African citizens' environmental action. Logically, believing that CC is caused by human action is the most important covariate of thinking that human action can mitigate its impacts. Far behind it, we find temperature anomalies, education level, CC risk perceptions and access to information, which also increase self-efficacy. Again, intolerant and authoritarian values are associated with reduced CCP. Households with fewer resources feel less empowered to fight against CC. Trust in institutions and speaking a western language increase self-efficacy. Finally, religiousness has a positive effect of about 1.5%.

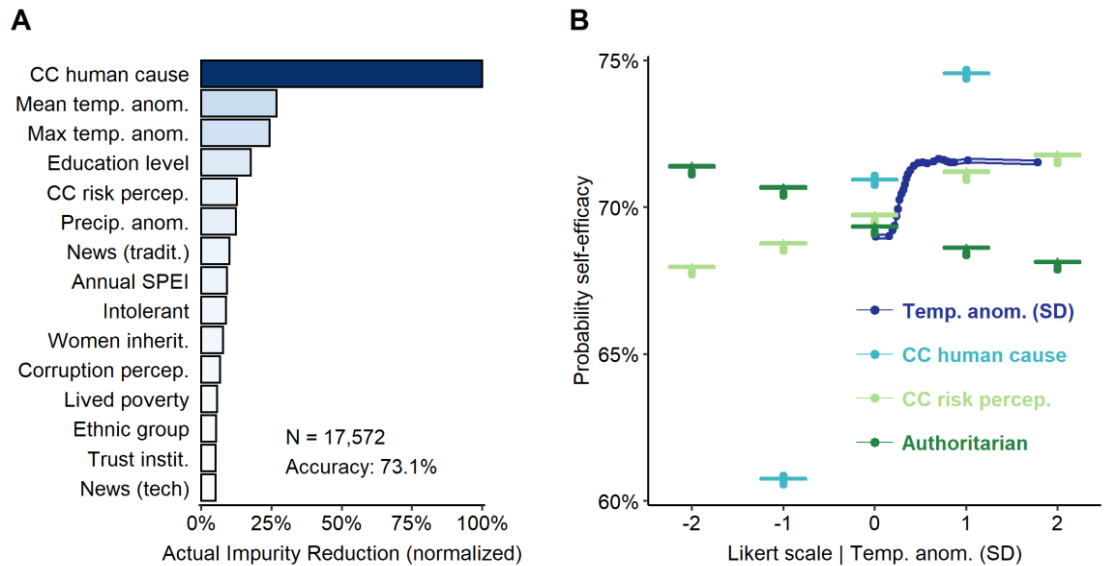

**Fig 5. Key predictors of environmental self-efficacy.**

(A) Top 15 predictors of self-efficacy. (B) Partial dependence plot of direct effects of mean temperature anomalies (*temp. anom.*), belief in human causation of CC (*CC human cause*), perceived risk from CC (*CC risk percep.*) and being favourable to one-man rule (*authoritarian*) on self-efficacy.

**Discussion**

These results show what shapes climate change perceptions (CCP) in Africa. Although each dimension of CCP has its unique set of predictors, some common patterns emerge from the analysis. First, the importance of perceived agriculture conditions stands out. Those individuals who perceive worsening agriculture conditions show higher awareness and perceived risk, more support for stopping CC and are more likely to believe it is caused by human action. The huge importance of the primary sector in terms of employment and export revenues makes agriculture a close and great concern to African citizens [49]. Thus, perceiving how CC is already affecting agriculture may reduce the psychological distance to CC. It is not a problem for "others" in space and time, it is happening here and now [50]. However, this relation also poses a challenge. CC has uneven impacts, and agriculture in some regions may benefit from changes in local climate [49]. Those who perceive those improvements are less likely to perceive CC and support or take environmental action. More efforts should be made to highlight the global nature of climate change and its overall negative impacts. These findings suggest environmental discourse in Africa could focus on the negative impacts of CC on agriculture to raise CCP, impulse individual adaptation and mobilise public support.

Second, attributing climate change to human activity increases risk perceptions, support for mitigation, and self-efficacy. If CC is *unnatural*, it is *extraordinary* and thus riskier, but also stoppable. Besides, attributing the cause of CC to human action might increase personal responsibility and, therefore, induce corrective responses [51,52]. This points to the convenience of spreading and highlighting the human origin of CC to impulse behavioural changes and mitigation strategies in Africa.

Risk perceptions are positively associated with self-efficacy and the need to stop CC. While some previous studies in the US and UK pointed to fatalism or climate despair [2,53,54]—where higher risks discourage self-efficacy and action support, the opposite seems to be true for Africa. This could be the result of motivated control —feeling more empowered to feel secure from a greater risk [55], or increased personal concern with CC [56]. Either way, framing CC as a critical risk will not discourage the African public, but it might encourage policy support and personal action [4].

Local weather conditions are among the most important predictors in all models, and on average they are more important than education, access to information or ideology. Previous research had found that *perceived* changes in local temperature were the most important predictor of CC risk perception in some African countries [11]. Building upon it, this study shows that *actual* long-term anomalies in temperature and rainfall at the second administrative level predict individual CCP across various dimensions. Attribute substitution and emotional salience may explain the importance of personal experience with local weather conditions for CCP [9,24]. Besides, qualitative evidence suggests that some communities in Africa understand climate change not as a global but a local phenomenon [35]. Therefore, local weather changes may be used to prime CC and encourage mitigation and adaptation, but the link between those local changes and the global nature of CC should be highlighted.

Information and education have great predictive power for being aware of CC and believing it has a human origin, the most analytical dimensions of CCP. On the other hand, they have less predictive power for more affective dimensions, such as risk perception or the need to stop CC. The limited emotional salience of CC information compared with personal experience or motivated reasoning might account for this divergence [9,20,21]. Nevertheless, the importance of information is contingent on language. Not speaking French, English or Portuguese hinders understanding climate terminology, which frequently lacks accurate translations to African languages [35].

Greater efforts should be made to translate to African languages the nature, causes and effects of CC.

Material conditions had previously been found to influence CCP, mainly by the mechanism known as the "finite pool of worry" [8]. According to it, worse material conditions limit CCP, as they create more urgent and pressing concerns to worry about. However, poverty has significant positive effects on risk perceptions and the need to stop CC across African countries. In contrast with the finite pool of worry hypothesis, poorer households are the most worried about the present and future effects of climate change. Their income and assets are the most vulnerable to climatic risks, so CC is an urgent concern for them [50,57].

Ideology has a significant impact on CCP in Africa. Authoritarian and intolerant ideologies are related to reduced CC awareness, belief in its human origin, risk perceptions, the need to stop CC and self-efficacy. These values have been consistent and negatively linked to CCP in other regions of the world [10]. Ideology influences what information people access, and how they process and assimilate it [15–19]. Authoritarians, through these mechanisms, disregard CC to justify their support for maintaining the status quo. These findings suggest it could be convenient to shape CC discourse to engage the authoritarian public. To do so, environmental discourse can frame policy and individual action as patriotism, innovation, or prosocial behaviour [58], and focus risk communication on the possible effects of CC on migration, security and public order

Political institutions play a crucial role in informing the public about CC and developing and implementing large-scale adaptation strategies. Therefore, trusting institutions has been found to reduce CC risk perceptions and increase policy support [6,59]. In Africa, trusting institutions shows a distinct pattern. It is associated with reduced CC awareness, belief in its human cause and risk perceptions. While the relation to risk perception is logical, the negative effect of trust on awareness and belief in human causation is surprising. Further research should address the possible mechanisms behind these findings.

Demographic variables such as gender, age or race have trivial overall importance. We only find an important gender gap for CC awareness. Women are less likely to be aware of climate change, as previous case studies in Africa had suggested [35,60]. Except for that, demographics are not among the most important predictors, in line with anterior research [10,11,13]. Moreover, although religion has been found to shape CCP in other settings [37,38], we find religion and religiousness to be mostly insignificant to predict CCP in Africa.

Urgent action is needed to limit the impacts of CC on ecosystems, economies, and political institutions in Africa. Knowing what shapes individual climate change perceptions across the continent contributes to the endeavour of raising awareness and policy support and encouraging self-efficacy and adaptive behaviour.

**Acknowledgements****Acknowledgements**